\documentclass[twocolumn]{aastex701}

\usepackage{xspace}
\usepackage[version=4]{mhchem}

\usepackage{tabularx}
\usepackage{array}

\newcommand{\firefly}{\texttt{FIREFLy}\xspace}

\newcommand{\exotedrf}{\texttt{exoTEDRF}\xspace}

\newcommand{\hst}{{\textit{HST}}\xspace}
\newcommand{\jwst}{{\textit{JWST}}\xspace}

\newcommand{\planetname}{{LHS~1140~b}\xspace}

\newcommand{\APL}{Johns Hopkins Applied Physics Laboratory, 11100 Johns Hopkins Rd, Laurel, MD 20723, USA}

\begin{document}

\title{No Helium Detected in LHS~1140~b from Four JWST NIRISS/SOSS Transits}

\author[0000-0002-9030-0132]{Katherine A. Bennett}
\affiliation{Department of Earth $\&$ Planetary Sciences, Johns Hopkins University, Baltimore, MD 21218, USA}
\email[show]{kbenne50@jhu.edu}  

\author[0000-0001-5097-9251]{Carlos Gascón}
\affiliation{Space Telescope Science Institute, 3700 San Martin Drive, Baltimore, MD 21218, USA}
\email{cgascon@stsci.edu}

\author[0000-0002-0746-1980]{Jacob Lustig-Yaeger}
\affiliation{\APL}
\email{Jacob.Lustig-Yaeger@jhuapl.edu}

\author[0000-0002-3263-2251]{Guangwei Fu}
\affiliation{Department of Physics $\&$  Astronomy, Johns Hopkins University, Baltimore, MD 21218, USA}
\email{guangweifu@gmail.com}

\author[0000-0001-6050-7645]{David K. Sing}
\affiliation{Department of Earth $\&$ Planetary Sciences, Johns Hopkins University, Baltimore, MD 21218, USA}
\affiliation{Department of Physics $\&$  Astronomy, Johns Hopkins University, Baltimore, MD 21218, USA}
\email{dsing@jhu.edu}

\author[0000-0002-7352-7941]{Kevin B. Stevenson}
\affiliation{\APL}
\email{Kevin.Stevenson@jhuapl.edu}

\author[0000-0002-2072-6541]{Jonathan Brande}
\affiliation{Department of Astronomy, University of Maryland, College Park, MD 20742, USA}
\email{jbrande@umd.edu}

\author[0000-0003-4157-832X]{Munazza K. Alam}
\affiliation{Space Telescope Science Institute, 3700 San Martin Drive, Baltimore, MD 21218, USA}
\email{malam@stsci.edu}

\author[0000-0003-3305-6281]{Jeff A. Valenti}
\affiliation{Space Telescope Science Institute, 3700 San Martin Drive, Baltimore, MD 21218, USA}
\email{valenti@stsci.edu}

\author[0000-0003-3204-8183]{Mercedes L\'opez-Morales}
\affiliation{Space Telescope Science Institute, 3700 San Martin Drive, Baltimore, MD 21218, USA}
\email{mlopez-morales@stsci.edu}

\author[0009-0004-4599-2477]{Sten J. Vermeiren}
\affiliation{School of Physics and Astronomy, University of St Andrews, North Haugh, St Andrews, KY16 9SS, UK}
\email{sv85@st-andrews.ac.uk}

\author[0000-0003-4241-7413]{Megan Weiner Mansfield}
\affiliation{Department of Astronomy, University of Maryland, College Park, MD 20742, USA}
\email{mwm@umd.edu}

\author[0000-0002-6721-3284]{Sarah E. Moran}
\affiliation{Department of Astronomy, University of Maryland, College Park, MD 20742, USA}
\email{semoran@umd.edu}

\author[0000-0001-7393-2368]{Kristin S. Sotzen}
\affiliation{\APL}
\email{Kristin.Sotzen@jhuapl.edu}

\author[0000-0003-2775-653X]{Jegug Ih}
\affiliation{Space Telescope Science Institute, 3700 San Martin Drive, Baltimore, MD 21218, USA}
\email{jih@stsci.edu}

\author[0000-0002-1046-025X]{Sarah Peacock} 
\affiliation{University of Maryland, Baltimore County, MD 21250, USA} 
\affiliation{NASA Goddard Space Flight Center, Greenbelt, MD 20771, USA}
\email{sarah.r.peacock@nasa.gov}

\begin{abstract}

In the effort to determine which low-mass exoplanets have atmospheres, LHS~1140~b remains one of the most favorable targets. Its large size (5.6 $\rm M_{\oplus}$ and 1.7 $\rm R_{\oplus}$) and relatively long orbital period (24.7 days) imply an atmosphere may be likely, and notably, recent interior models favor either a hydrogen-dominated ``mini-Neptune" or a ``water world" over a true terrestrial planet. Another possibility is that it has a helium-rich atmosphere. This hypothesis is supported by recent ground-based observations that detected the metastable helium triplet during transit. These observations indicated there may be current helium escape from the planet's upper atmosphere, yet the signal was not detected during a subsequent observation, suggesting time-variable escape. Here we present four observations of LHS~1140~b with JWST NIRISS/SOSS, which covers the metastable helium triplet, obtained between 2023 and 2026. These observations span the epoch of the ground-based measurements, and although none were contemporaneous with the ground-based transits, all four are sensitive to helium absorption at the previously reported level. However, we detect no helium absorption in any visit. We reject the best-fit ground-based model at $>3\sigma$ in each visit, and find no clear trend in mass-loss with time. Our results suggest the reported ground-based detection may be spurious, although variability cannot be excluded if detectable helium absorption occurs in $\lesssim50\%$ of transits. The nature of LHS~1140~b thus remains a mystery until future transmission and emission analyses are complete.  

\end{abstract}


\section{Introduction} \label{sec:intro} 

The quest to understand the nature of low-mass exoplanets continues to both inspire and plague exoplanet astronomers. The simple yet central question --- \textit{which of these exoplanets have atmospheres?} --- has proven quite challenging to answer (e.g., \citealt{Lim2023,Lustig-Yaeger2023, Moran2023, Zieba2023, Alderson2024,AdamsRedai2025,Espinoza2025, Glidden2025,Radica2025, Wachiraphan2025}). Still, it is a question worth dedicating time to, as mapping the boundary between planets with and without atmospheres -- the so-called ``cosmic shoreline" \citep{Zahnle2017} -- is the first step toward quantifying the prevalence of possibly habitable planets in the nearby Galaxy. Today, we parameterize this hypothetical boundary by considering a planet's escape velocity and cumulative high-energy incident stellar flux: the higher the planet's escape velocity and the farther it is from its host star, the more likely it may be that it retains an atmosphere over geologic timescales. Of course, this framework is further complicated by the fact that the planets that are most observationally accessible - tidally-locked planets orbiting close to M dwarfs - may also be the least likely to retain atmospheres due to the high X-ray and ultraviolet (XUV) flux and activity levels of their stellar hosts (e.g., \citealt{Khodachenko2007,Lammer2007,Owen2012, Peacock2019,vanLooveren2024,Pass2025}).

However, under the assumption that there \textit{are} some M dwarf planets able to retain atmospheres, it follows that we should focus our search on the largest and least XUV-irradiated planets known. The single best target found so far is the nearby, $\rm 5.60\pm0.19\;M_{\oplus}$ and $\rm 1.730\pm0.025\;R_{\oplus}$ planet \planetname, orbiting its 3096~K host in 24.7 days, giving it a zero-Bond-albedo equilibrium temperature of just 226~K \citep{Cadieux2024b}. One of the first temperate small-mass planets discovered \citep{Dittmann2017b}, this planet remains one of the community's highest priority targets. Indeed, it is a planned target of the 500-hour Rocky Worlds Director's Discretionary Time Program, which aims to search for atmospheres around optimal yet observationally intensive targets. Initial reconnaissance transmission spectroscopy of \planetname with the \textit{Hubble Space Telescope}'s (\hst) WFC3/G141 observing mode tentatively revealed an atmospheric water feature, but the authors concluded that the low S/N feature could also be due to stellar contamination and looked toward promising future \jwst observations to elucidate the nature of this planet \citep{Edwards2021}.

However, interpreting \planetname's spectrum has been further complicated in recent years by updated mass and radius estimates for this planet, suggesting \planetname is not actually a terrestrial planet in the traditional sense. That is, it may not be a rocky body surrounded by a thin, high mean molecular weight atmosphere. While initial constraints found that \planetname was consistent with an Earth-like structure \citep{Ment2019, LilliBox2020}, a more recent interior structure modeling analysis used updated mass and radius constraints to conclude the planet is either a ``mini-Neptune" (a rocky body surrounded by a low mean-molecular weight, more extended atmosphere) or a ``water world" (a body made of rock alongside a substantial portion of water, in this case, 9-19\% by mass) \citep{Cadieux2024b}. This makes sense in light of demographics studies that have found most planets $\rm > 1.6\;R_{\oplus}$ are unlikely to be rocky \citep{Roger2015, Cloutier2020}. 

\jwst transmission spectroscopy using NIRSpec/G235H and G395H favored the water world interpretation, as they report a featureless spectrum inconsistent with an $\rm H_2$-dominated atmosphere, which should produce large features \citep{Damiano2024}. They argue instead that \planetname likely has a high mean molecular weight ``secondary" atmosphere. Clouds, which notably could also explain a featureless spectrum, should not form and persist as an opaque layer at the high altitudes probed by transmission spectroscopy \citep{Robinson2014, Moran2018, Cadieux2024}, and hazes are unexpected at these altitudes for a planet with as low an equilibrium temperature as \planetname \citep{Yu2021}. Using \jwst's NIRISS/SOSS, \cite{Cadieux2024} also favor a high mean-molecular weight atmosphere + water world interpretation, tentatively discerning a Rayleigh-scattering signal from an $\rm N_2$-dominated atmosphere hidden underneath stellar contamination features. 

A third possibility is that \planetname instead has a helium-dominated atmosphere driven by mass fractionation during atmospheric escape \citep{Hu2015, Malsky2020, Malsky2023, Cherubim2024,Lammer2025, Cherubim2025,deWit2026}. Assuming a planet starts off with a bulk H/He envelope, hydrogen may preferentially escape due to atmospheric diffusive separation if the hydrodynamic escape rate is comparable to the diffusion-limited escape rate \citep{Hu2015}, gradually leading to hydrogen-depleted and helium-enriched atmospheres over billions of years. \citet{Cherubim2026} recently reported a detection of the metastable helium triplet in \planetname via a ground-based detection during a transit of the planet using the WINERED spectrograph on the Clay/Magellan II telescope, supporting this hypothesis. The metastable helium triplet is one of the most robust indicators of atmospheric escape, leading \cite{Cherubim2026} to infer that \planetname has a helium-dominated upper atmosphere currently undergoing escape, with high mean molecular weight species trapped deeper in the atmosphere, below the level of the transit photosphere. They note that this picture of a changing mean molecular weight with altitude is consistent with the featureless spectrum from \cite{Damiano2024}, as a clear helium-dominated atmosphere all the way down would have features discernible with \jwst.

If this signal proves robust, it would be a large step forward in our attempt to elucidate the nature of \planetname and map the cosmic shoreline across low-mass planets. However, \cite{Cherubim2026} only detected helium once in 2024, with a second nondetection reported in 2025, which they attribute to time-variable escape. More observations should therefore further illuminate whether 1) this 2024 signal is representative of the typical state of \planetname, 2) the escape is more stochastic and erratic, or 3) the 2024 signal was erroneous. 

Executed \jwst NIRISS/SOSS observations offer us a chance to place this ground-based detection in context. While SOSS offers a much lower resolution than ground-based data, its precision allows us to detect signals even when they are diluted \citep{Fu2022,Ahrer2025, Allart2025,Fournier2025,Krishnamurthy2026, Mukherjee2026}. \cite{Cherubim2026} detected a 1.24\% absorption depth, which should translate into a $\sim$600~ppm feature at pixel resolution with SOSS. (We detail this calculation in Section \ref{sec:analysis_model_convolution}.) 

In this paper, we compare the \cite{Cherubim2026} ground-based results to four transits of \planetname taken with NIRISS/SOSS between 2023--2026, described in Sections \ref{sec:observations} and \ref{sec:reductions}. We do not detect helium in any of our four transits, and we demonstrate that the ground-based detection is significantly discrepant with our data (Section \ref{sec:analysis}). Finally, we discuss the implications of our findings and offer a path forward in Section \ref{sec:discussion}.

\section{Observations} \label{sec:observations}

We analyze data from four transits of \planetname taken with \jwst NIRISS/SOSS. Two transits (Visits 1 and 2, hereafter) are archival, obtained on 2023~December~01 and 2023~December~26 as part of the program DD~6543 (PIs: C. Cadieux \& R. Doyon), and were originally published by \cite{Cadieux2024}. Two transits (Visits 3 and 4, hereafter) are new, obtained as part of the program GO~7073 (PIs: J. Lustig-Yaeger \& K. Stevenson). These were taken on 2025~August~10 and 2026~July~23. 

All four observations used the SUBSTRIP256 subarray with the NISRAPID readout pattern. Visits 1 and 2 used three groups per integration and 949 total integrations (5.79 hours of observation), while Visits 3 and 4 used five groups per integration, for 547 total integrations (5.00 hours of observation). The V3 position angles are: 54.4\textdegree\xspace for Visit~1, 67.4\textdegree\xspace for Visit~2, 262.3\textdegree\xspace for Visit~3, and 257.6\textdegree\xspace for Visit~4. LHS~1140 itself was used for target acquisition, which was done with 3 groups for Visits 1/2 and 11 groups for Visits 3/4 utilizing the SOSSFAINT mode and the F480M filter with the NISRAPID readout pattern. An F277W exposure was also taken for all four transits to assess zeroth order background contaminants shortward of $\rm 2.4\;\mu m$. This 10-integration exposure was taken with 3 groups for Visits 1/2 and 6 groups for Visits 3/4. No zeroth order background contaminants overlap the spectrum in any visit.

\section{Data Reduction} \label{sec:reductions}

For this study, we focus on the narrow wavelength range surrounding the 1.0833 $\rm \mu m$ metastable helium triplet (specifically, $\rm 1.07-1.095\;\mu m$), leaving the analysis of the complete $\rm 0.6-2.8\;\mu m$ transmission spectra of \planetname for future work. Because the metastable helium triplet feature is narrower than a single SOSS pixel, we produce light curves at the column level to maximize the S/N of any helium feature. We reduce all four transits with the \firefly pipeline \citep{Rustamkulov2022, Rustamkulov2023,Wang2026}, detailed below, and confirm the \firefly pipeline replicates the transmission spectrum presented by \cite{Cadieux2024} for Visits 1 and 2. For our two new transits, Visits 3 and 4, we also reduce the data with \exotedrf \citep{Radica2023, feinstein2023, Radica2024} to confirm \firefly's results, as is standard for \jwst studies. A description of the \exotedrf pipeline and a comparison between the reductions can be found in \autoref{sec:appendix_reduction_comparison}.

\subsection{\firefly} \label{sec:firefly}

\firefly takes the \texttt{uncal.fits} downloaded from MAST and applies a modified version of Stages 1 and 2 from the \texttt{jwst} pipeline to extract the up-the-ramp reads prior to custom cleaning and light-curve fitting. We use a jump rejection threshold of 6$\sigma$ for all visits, and apply a temporal group-level 1/f correction after subtracting STScI's publicly available background model scaled to the flux of our measured background, with a separate scaling factor applied to the left and right side of the jump in background at x$\sim750$ \citep{Lim2023, Fournier-Tondreau2024, Wang2026}. We customize which regions of the detector are used to measure the background by visit (and therefore location of first and second order contaminants). The background is then added back in prior to up-the-ramp fitting, as it represents true astrophysical (i.e., not readout) counts. See \citealt{Wang2026} for details on these group-level steps. We also confirm our up-the-ramp reads are not impacted by nonlinearity. 

Next, we clean the 2D integrations spatially and temporally, noting that we apply a conservative ($\sigma>10$) temporal cleaning threshold to avoid the appearance of correlated noise in the column-level transmission spectra. We then permanently subtract our scaled background model and apply a final 1/f correction at the integration-level. We interpolate the spectra onto a common grid using the measured intra-pixel position shifts in both the spatial (y) and spectral (x) direction. To extract the 1D stellar spectra, we use box extraction with an aperture full-width of 36 pixels along the trace, which is measured using \texttt{pastasoss} \citep{Baines2023a, Baines2023b}. We note that the precise width of the aperture, while improving the scatter in the white light curve, makes effectively no difference in the scatter of the low S/N column-level light curves. As discussed in \cite{Cadieux2024}, Visit 1 was executed with blind pointing, causing the trace and thus wavelength solution to be displaced from their nominal positions. \cite{Cadieux2024} corrected this by displacing the nominal trace/wavelength solution by $\rm \Delta x=-157$ pixels and $\rm \Delta y=-12$ pixels, which they determine by cross-correlating the stellar trace with a PHOENIX stellar model. However, we now have the benefit of three other LHS~1140 SOSS observations executed without this erroneous offset, and we use these observations to empirically determine that $\rm \Delta x=-167$ pixels and $\rm \Delta y=-14$ pixels are the optimal offsets. The SOSS trace position depends on the pupil wheel position \citep{Baines2023b} in addition to the x-direction offset, meaning that the order of the translation and rotation operations when determining the position of the spectral trace could affect the final wavelength solution (as rotations and translations do not commute). While \texttt{FIREFLy} does not take this into account (passing the pupil wheel position into \texttt{pastasoss} first, then translating the trace model by the x-offset), we note that this only introduces a $\sim0.2$~pixel offset in the position of our desired helium line in Visit 1, as compared to first translating the modeled trace and then applying the pupil wheel rotation. This falls within the uncertainty in the \texttt{pastasoss} wavelength solution \citep{Baines2023a} and thus does not impact our results. 

We then move to light curve fitting, modeling an transit plus systematics model of the white light curve to derive system parameters: semi-major axis normalized to the number of stellar radii ($a/R_s$), impact parameter ($b$), mid-transit time ($T_0$), quadratic limb darkening ($u_+$ and $u_-$, \citealt{Brown2001b}), planet-to-star radius ratio $(R_p/R_s$), and baseline flux. We take the period to be 24.73691 days \citep{Edwards2021} and fix the eccentricity and argument of periastron to zero. Because we are not concerned with the white light transit depths for this analysis (we normalize our spectrum to one), we do not insist on consistent system parameters between visits. We use the BPICS criterion \citep{Thorngren2026} to determine the best-fit systematics model. For Visits 1 and 4, we use a linear model. For Visit 2, we use a linear term along with the square of the x- and y- position shifts, and for Visit 3, we use a 4th order polynomial. We use \texttt{emcee} \citep{emcee2013} to fit the transit plus systematics model, with 2000 burn-ins and 6000 steps. As shown by \cite{Cadieux2024}, in Visit 2, \planetname's transit is interrupted by a serendipitous transit of planet c (this can be seen in \autoref{fig:light_curves}), so we also fit for planet c's radius and system parameters in this visit. Finally, there are some possible minor spot crossings visible in Visit 4's light curve. While spots do impart a wavelength-dependent change in transit depth, they do so over a much broader wavelength range than what we are interested in for this study. We thus ignore the possible spots in our fits, noting that the \exotedrf reduction in Section \ref{sec:exotedrf} tests both masking and ignoring them, and the shape of the resulting transmission spectrum is not impacted. 

For the spectroscopic fits, we fit each column independently using \texttt{lmfit} \citep{Newville2014}, fitting only for flux and $R_p/R_s$, keeping all other astrophysical and systematics parameters fixed to their white light curve values. This includes the linear systematics term, which is constant within this narrow wavelength range, though it does have a wavelength-dependency at larger wavelength scales. We also fix limb darkening to a PHOENIX model \citep{Husser2013} using \texttt{exoTIC-LD} \citep{Grant2024} assuming $T_{\rm eff}$=3096, log~g=5.041 (cgs), and [M/H]=--0.15 for the host star \citep{Cadieux2024b}, rescaling the intensity profiles to set $\mu=0.0$ at the edge of the photosphere and only using $\mu>0.2$ to calculate the limb darkening coefficients. Finally, we fix planet c's radius in Visit 2 to its white light curve value, as planet c's radius is not expected to change over such a short wavelength regime (indeed, it may be a bare rock and so may not change at all; \citealt{Fortune2025, Rochon2026}), and metastable helium was not detected in planet c \citep{Cherubim2026}. Our transmission spectra for all four visits are shown in \autoref{fig:trans_spectra_4visits}.

\section{Analysis} \label{sec:analysis}

\subsection{No Excess Absorption in Helium Light Curves}

\begin{figure*}
    \centering
    \includegraphics[width=1\linewidth]{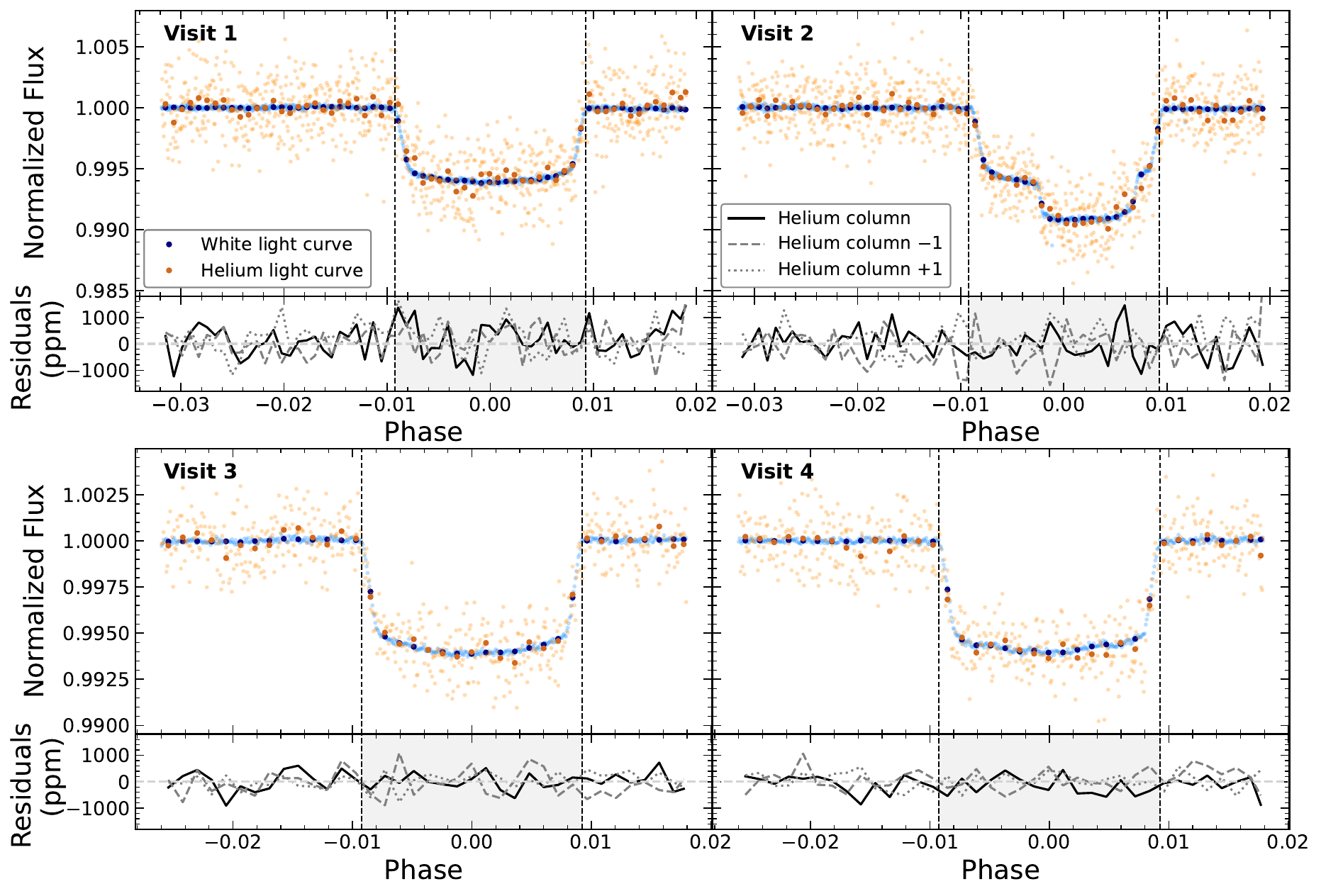}
    \vspace{-0.6cm}
    \caption{Column-level light curves containing the 1.0833~$\rm \mu m$ helium triplet (orange) compared to the white light curves (blue) for all four visits. Unbinned and binned points are shown in light and darker shades, respectively. We note Visits 1 and 2 display higher scatter relative to Visits 3 and 4 because they have fewer groups per integration; if all visits are binned to the same time sampling, the scatter is similar. Bottom panels show residuals between the binned helium light curves and the white light curves (black solid lines). Also shown are residuals between the two adjacent columns on either side of the helium column and the white light curves (gray dotted and dashed lines), examined in case some of the helium signature has leaked into adjacent columns. There is no evidence for excess helium absorption in any of the four light curves.}
    \label{fig:light_curves}
\end{figure*}

To examine if there is evidence for helium escape from the atmosphere of \planetname, we first compare the column-level light curves expected to contain the helium triplet absorption to the white light curves for each visit, as shown in \autoref{fig:light_curves}. All three lines of the helium triplet should fall within a single SOSS pixel. Visually, there is no excess absorption during transit in the helium light curve in any of the four visits, nor is there evidence for pre- or post-transit helium tails. We confirm this by examining the residuals between the helium and white light curve (black solid lines in the bottom diagrams of each panel in \autoref{fig:light_curves}), which are structureless. We also examine the column-level light curves adjacent to that expected to contain helium absorption, in case the signature is also present in adjacent columns due to Nyquist sampling \citep[see][]{Albert2023}. The residuals between the light curves of the columns to the left and right ($-1$ and $+1$, represented as dashed and dotted lines, respectively, in the residual panels in \autoref{fig:light_curves})
of the column expected to contain helium absorption and the white light curve are also structureless. Thus, there is no evidence for any excess absorption from the light curves alone. 

\subsection{Limits on Helium Absorption Depth}

We show the four NIRISS/SOSS transmission spectra in \autoref{fig:trans_spectra_4visits}. No helium absorption is visible by eye in any of the four visits. In order to place our nondetections in context, we calculate the $2\sigma$ upper limits on the helium absorption depths at the pixel-resolution of our SOSS data. We report these limits in \autoref{tab:limits}. We only consider absorption features with line profiles that have the fixed morphology of the \cite{Cherubim2026} best-fit model, convolved to the resolution of SOSS. As absorption depth is resolution-dependent, we convert these upper limits into the equivalent upper limits at the WINERED resolution to facilitate comparison with the ground-based results; these values are also reported in \autoref{tab:limits}. Our $2\sigma$ upper limits at the ground-based resolution are similar to the 0.6\% upper limit reported in the 2025 nondetection from \cite{Cherubim2026}.

\begin{deluxetable}{c|ccc}
\tablewidth{0pt}
\tablecaption{$2\sigma$ upper limits on helium absorption depth, reported at both NIRISS/SOSS and the ground-based WINERED resolutions, and mass-loss rate. \label{tab:limits}}
\tablehead{
    \colhead{Visit} & 
    \colhead{Depth} & 
    \colhead{Depth} &
    \colhead{$\dot{M}$} \\[-2mm]
    \colhead{} & 
    \colhead{\textit{SOSS resolution}} & 
    \colhead{\textit{WINERED resolution}} & 
    \colhead{($\rm g\;s^{-1}$)}
}
\startdata
1 & 0.03\% & 0.45\% &$1.4\times10^7$\\
2 & 0.02\% & 0.41\% &$5.4\times10^6$\\
3 & 0.04\% & 0.61\% &$3.1\times10^7$\\
4 & 0.03\% & 0.44\% &$8.0\times10^6$\\
\enddata
\tablecomments{$\dot{M}$ is calculated using \texttt{pwinds} assuming all parameters equivalent to those presented in \cite{Cherubim2026}.}
\end{deluxetable}

\subsection{Ground-Based Model Discrepant with All Four NIRISS/SOSS Transmission Spectra} \label{sec:analysis_model_convolution}

\begin{figure*}
    \centering
    \includegraphics[width=1\linewidth]{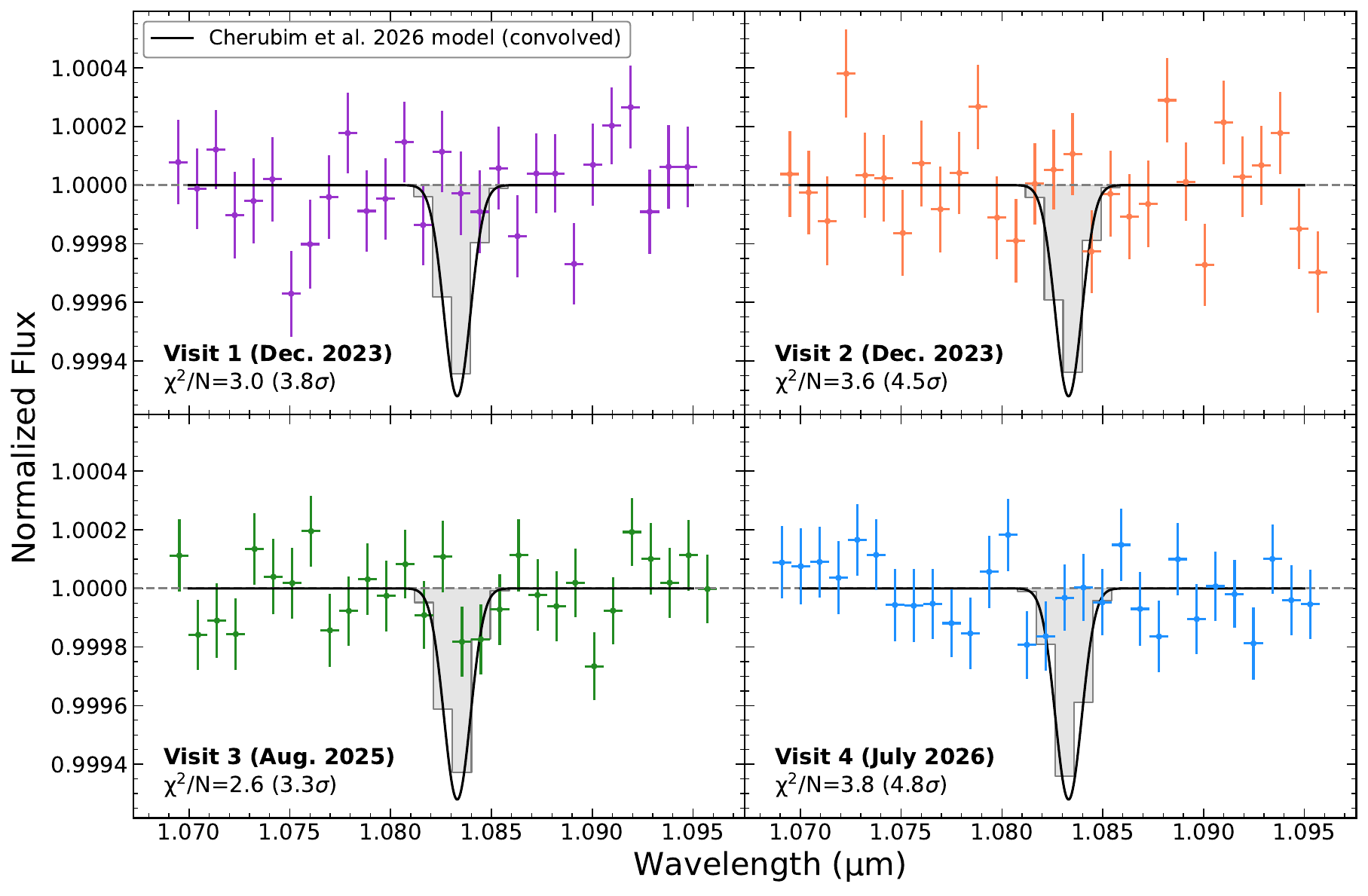}
    \vspace{-0.6cm}
    \caption{The NIRISS/SOSS transmission spectra of \planetname centered around the 1.0833~$\rm \mu m$ metastable helium triplet from four visits spanning 2023 -- 2026. Overplotted is the best-fit model reported by \cite{Cherubim2026} from the September~2024 ground-based observation, convolved to the resolution of SOSS (black line) and binned to the pixel scale (gray histogram). Every one of the four spectra is discrepant from the ground-based data by $>3\sigma$. If the reported 1.24\% ground-based absorption depth were consistent over time, we would detect it in the NIRISS/SOSS data.}
    \label{fig:trans_spectra_4visits}
\end{figure*}

To compare with the reported ground-based detection, we first determine if the excess absorption observed from the ground in 2024 would be detectable with SOSS. We consider the best-fit model reported by \cite{Cherubim2026}. The authors used \texttt{pwinds} \citep{dosSantos2022}, which computes the metastable helium population as a function of altitude assuming a bulk H/He atmosphere and an isothermal Parker wind. Inputs include the planetary system parameters, as well as stellar XUV flux, the atmospheric hydrogen number fraction relative to helium (H:He), mass-loss rate ($\dot{M}$), temperature of the outflow ($T_{\rm wind}$), and line-of-sight velocity of the outflow ($v_{\rm wind}$). Inversely, the strength of the helium feature can be used to back out $\dot{M}$, $T_{\rm wind}$, H:He, and $v_{\rm wind}$. This is the approach taken by \cite{Cherubim2026}, who found that, using the XUV flux of GJ~1132 as a proxy for that of LHS~1140, $\dot{M}=2.03^{+0.67}_{-0.58}\times10^8{\rm \;g\;s^{-1}}$, $T_{\rm wind}=5160^{+46}_{-50}$~K, H:He = $1.01^{+0.85}_{-0.50}\times10^{-3}$, and $v_{\rm wind}=2.26^{+0.33}_{-0.30}{\rm \;km\;s^{-1}}$ provided the best fit for their data. 

We then convolve this best-fit model to the resolution of SOSS to determine if the same signature would be detectable with our low-resolution observing mode. As the convolution is critical to our subsequent interpretation, it is worth sharing a few details here on how we approached this. We convolve the ground-based model sampled at the resolution of the ground-based data with an idealized Gaussian kernel representing the SOSS PSF using the \texttt{scipy} function \texttt{gaussian\_filter1d}. We assume the kernel is directly centered on each SOSS pixel. To estimate the $1\sigma$ width of the Gaussian SOSS PSF, we calculate the FWHM of the PSF at the helium triplet assuming $R\approx700$ (the published resolution from JDox), finding $\rm FWHM\approx15.5\; \AA$, then convert this into $\sigma$.

We also explored a more empirical approach, in which we utilize the publicly available SOSS kernel measured during on-the-ground cryo-vacuum testing prior to \jwst's launch.\footnote{This information can be found in the file \texttt{jwst\_niriss\_speckernel\_0004.fits} from the \jwst Calibration Reference System website: \url{https://jwst-crds.stsci.edu/}.} However, we choose to present the idealized Gaussian kernel in this work for several reasons. First and foremost, the empirically measured PSF is slightly asymmetric, but the directionality (toward increasing or decreasing wavelengths) of this asymmetry cannot be determined from the \texttt{fits} file alone. Secondly, many previous studies have approximated the SOSS PSF as a Gaussian function \citep{Fu2022,
Ahrer2025, Fournier2025}. Finally, the choice between the Gaussian and empirical PSF does not impact our results, as the two kernels display differences that would be very challenging to discern at the pixel scale of SOSS. 

Thus, regardless of choice of kernel used for convolution, our data are discrepant with the ground-based model, as shown in \autoref{fig:trans_spectra_4visits}. As can be seen in the figure, if the 1.24\% absorption feature measured with WINERED in 2024 was consistent over the years, we would have been able to confidently detect it with SOSS at each epoch. However, we do not see helium absorption across any of our four visits. We calculate $\chi^2$ between our data and the convolved model, normalized by the number of data points ($\chi^2$/N), focusing just on the region around the helium feature, $\rm1.077-1.090\;\mu m$. Converting this to $\sigma$ equivalences indicates that each SOSS transit rejects the ground-based best-fit model to $>3\sigma$. Combining all four transits into one weighted mean transmission spectrum, we reject the ground-based model to $9.9\sigma$.

\subsection{Upper Limits on Possible Mass Loss Rates}

\begin{figure}
    \centering
    \includegraphics[width=1\linewidth]{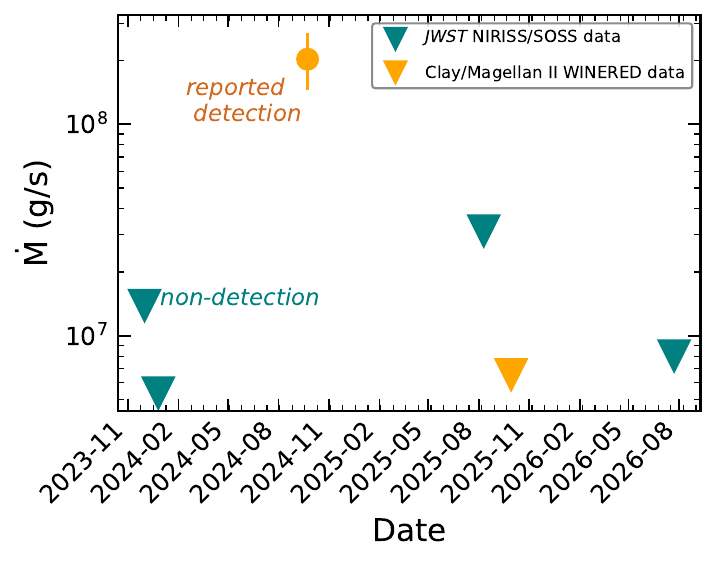}
    \vspace{-0.6cm}
    \caption{Mass-loss rates (single point with $1\sigma$ error bars) and $2\sigma$ upper limits (triangles) derived from \texttt{pwinds} for all six visits: two from ground-based WINERED observations (orange) and four from \jwst SOSS (teal). Orange values are taken from \cite{Cherubim2026} and teal points are calculated assuming the same input parameters as \cite{Cherubim2026}, varying only the mass-loss rate. The single reported detection from 2024 is a notable outlier, and the six visits together reveal that there is no apparent trend over time.}
    \label{fig:mass_loss_rates}
\end{figure}

Though we do not find any evidence for helium absorption, we can use our data to place upper limits on any possible mass-loss rate occurring on \planetname in order to put \cite{Cherubim2026}'s findings in context. The authors offer variable helium escape as an explanation for their 2025 nondetection, possibly due to varying XUV stellar flux or varying upper atmosphere temperatures ($T_{\rm wind}$). By combining our findings with the ground-based results, we can leverage findings from six total observations spanning nearly three years to explore whether variable escape is a viable explanation.

From a measured absorption depth of 1.24\% with WINERED in 2024, those authors found a mass-loss rate of $\dot{M}=2.03^{+0.67}_{-0.58}\times10^8{\rm \;g\;s^{-1}}$. 
They also placed a conservative upper limit of 0.6\% absorption depth from their 2025 nondetection, which, assuming the same parameters as their best-fit model, translates into an upper limit of $\dot{M}=6.6\times10^6{\rm \;g\;s^{-1}}$.

To compare findings, we use this same approach and compare \texttt{pwinds} models with our data to determine what mass-loss rates are consistent with our nondetections.
We vary only mass-loss rate and keep all other input parameters fixed to the values from \cite{Cherubim2026}, which are available on Zenodo\footnote{\url{https://zenodo.org/records/20723095}}. Following \cite{Cherubim2026}, we use the XUV flux of GJ~1132 as a proxy for LHS~1140. We determine which values of $\dot{M}$ we can reject to at least $2\sigma$. Our upper limits for each visit are reported in \autoref{tab:limits}. These limits, along with the limits and reported detection from the ground-based data, are also shown in \autoref{fig:mass_loss_rates} as a function of time. With all six observations (two with WINERED and four with SOSS), there is no clear $\dot{M}$ trend over time. The singular reported detection (on the order of $\rm 10^8\;g\;s^{-1}$) in 2024 stands out as an outlier, with all the nondetections reporting consistent upper limits on the order of $\rm 10^6-10^7\;g\;s^{-1}$.

We note several caveats to this analysis. For one, at mass-loss rates approaching $\rm 10^9\;g\;s^{-1}$, the helium feature begins \textit{decreasing} in magnitude due to hydrogen self-shielding of XUV radiation that decreases the population of the metastable state \citep{Oklopcic2019}. This means that technically, our nondetections are also consistent with very high ($\rm> 10^9\;g\;s^{-1}$) mass-loss rates. However, given the age of the system ($>5$~Gyr old), it is unlikely that such a high mass-loss rate could feasibly be maintained over geologic timescales for such a low-mass planet. 
Additionally, we explore varying only a single parameter - $\dot{M}$ - within the \texttt{pwinds} framework, though in reality the amplitude of the helium feature is driven by a complex interplay between mass-loss rate, temperature, H:He number fraction, and stellar XUV flux, among other parameters. Without resolving the helium line profile, many of these parameters are degenerate \citep{dosSantos2022}. Furthermore, factors such as the stellar EUV flux (either time variable or single epoch) and precise temperature of the upper-atmosphere are model-dependent and not well-characterized. All of this complicates our ability to directly infer mass-loss rates from the amplitude of the helium feature. However, it is beyond the scope of this paper to explore the physical intricacies of modeling metastable helium (e.g., \citealt{Oklopcic2019,Lampon2020,dosSantos2022}). We thus caution the reader to not interpret \autoref{fig:mass_loss_rates} as conveying strict boundaries on possible mass-loss rates, but instead demonstrating the relative mass-loss rates allowed by different data sets given uniform modeling assumptions. Importantly, compared to the helium escape observed in 2024, \autoref{fig:mass_loss_rates} shows no other instances of similar levels of atmospheric escape in all other observations collected over a period of three years. 

\section{Discussion and Conclusion} \label{sec:discussion}

Given our current set of six total observations, the single 2024 ground-based reported helium detection appears to be an outlier. If this observation is a true detection of planetary helium escape, escape may be more stochastic or rare than inferred by \cite{Cherubim2026}. Given our SOSS nondetections, we can make a few statements about the probability of various outcomes using a variable amplitude spectral template for the helium signal. The probability of four nondetections with SOSS under the assumption of a \textit{persistent} helium signal at the level observed by \cite{Cherubim2026} in 2024 is vanishingly small. However, time variability in the helium signal remains a possibility, albeit a less likely scenario following four SOSS nondetections. If \planetname exhibited \cite{Cherubim2026}'s reported 2024 signal during only a random fraction of the time, $f$, then for a duty-cycle of $f=50\%$, we would expect four nondetections with SOSS only ${\sim}6.3\%$ of the time (or, inversely, $f< 53\%$ at $2 \sigma$). While this is still uncommon, it does emphasize that time-variability in the signal cannot be as readily rejected by the current collection of nondetections as the 2024 reported signal in each SOSS visit.    

One more SOSS transit of \planetname is scheduled for November 2026 as part of our GO 7073 program, which will further shed light on whether escaping helium could be present. It is also challenging to quantify the timescale of variability, if it exists, with only a single detection and five nondetections over a three year timescale. Dedicated monitoring of LHS~1140~b transits could provide greater statistical leverage on the rarity of the 2024 detection, its false positive likelihood, and the timescale of any variability. 

While our helium absorption nondetections cast doubt on the helium-dominated upper atmosphere interpretation of \planetname, it does not constrain the relative likelihood of \planetname being a mini-Neptune versus a water world. We leave this to future, panchromatic analyses of \planetname's transmission spectrum, coupled with emission photometry results from the Rocky Worlds DDT program for this target. For now, we simply conclude by reiterating that our NIRISS/SOSS observations of \planetname are significantly at odds with the ground-based detection of metastable helium. Increasing the number of transits can resolve this controversy, but with six currently available observations, it is now incontrovertible that the previously detected helium absorption is not persistent, and if that signal is time-varying, then it recurs less than half of the time. At this point, it is unlikely though still possible that the ground-based signal represented a true escaping atmosphere from the ever-mysterious LHS 1140 b.

\begin{acknowledgments}
Katie Bennett thanks Patrick McCreery for engaging discussions on helium and stellar variability, and Le-Chris Wang for helpful discussions regarding correlated noise. 
This work is based on observations made with the NASA/ESA/CSA James Webb Space Telescope. The Claude Opus 4.5 large language model was used to improve plot aesthetics. The data were obtained from the Mikulski Archive for Space Telescopes at the Space Telescope Science Institute (STScI), which is operated by the Association of Universities for Research in Astronomy, Inc., under NASA contract NAS 5-03127 for JWST. These observations are associated with program \#6543 and \#7073. Support for program \#7073 was provided by NASA through a grant from STScI. 
This material is based in part upon work performed as part of the CHAMPs (Consortium on Habitability and Atmospheres of M-dwarf Planets) team, supported by the National Aeronautics and Space Administration (NASA) under Grant No. 80NSSC23K1399 issued through the Interdisciplinary Consortia for Astrobiology Research (ICAR) program. 
M. L-M. is supported by individual research time under NASA contracts NAS5-26555 and NAS5-03127 to the Associated Universities for Research in Astronomy for the operation of the Hubble Space Telescope and James Webb Space Telescope Science Operations Centers at STScI. S.P. acknowledges support from NASA under award number 80GSFC24M0006.
\end{acknowledgments}

\section*{Data Availability}
All data presented in this article were obtained from the Mikulski Archive for Space Telescopes (MAST) at the Space Telescope Science Institute (STScI). The four transit observations can be accessed via \dataset[doi:10.17909/yb6q-yk46]{https://doi.org/10.17909/yb6q-yk46}. \\
Data products are available on Zenodo: \url{https://zenodo.org/doi/10.5281/zenodo.21921754}. 

\facilities{ADS, JWST(NIRISS)}

\software{astropy \citep{astropy:2013, astropy:2018, astropy:2022}, \texttt{batman} \citep{Kreidberg2015}, 
Claude \citep{Anthropic2025_claude},
\texttt{emcee} \citep{emcee2013}, \texttt{exoTEDRF} \citep{Radica2023, feinstein2023, Radica2024} 
\texttt{ExoTIC-LD} \citep{Grant2022, Grant2024}, \firefly \citep{Rustamkulov2022, Rustamkulov2023}, 
\texttt{jwst} \citep{JWST_pipeline}, 
\texttt{lacosmic} \citep{vanDokkum2001}, 
\texttt{lmfit} \citep{Newville2014}, 
matplotlib \citep{Hunter2007}, 
numpy \citep{Harris2020}, 
pandas \citep{McKinney2010}, 
\texttt{PHOENIX} \citep{Allard2012, Husser2013}, 
\texttt{pwinds} \citep{dosSantos2022},
scipy \citep{Virtanen2020}
}

\appendix

\section{Comparison with the \exotedrf Pipeline} \label{sec:appendix_reduction_comparison}

\subsection{\exotedrf} \label{sec:exotedrf}
As an independent analysis, we reduce our newly acquired NIRISS/SOSS observations with the \exotedrf pipeline \citep{Radica2023, feinstein2023, Radica2024}. We reduce both visits following a similar set of steps and assumptions as the \firefly pipeline. We start our reduction from the \texttt{uncal.fits} files extracted from MAST and run the standard \exotedrf Stage 1 steps, which include data quality initialization, superbias subtraction, non-linearity correction and ramp fitting. We also correct for the 1/f noise at the group level, using the \texttt{scale-achromatic} method. During the 1/f correction, we mask the target's 1st, 2nd and 3rd order traces, as well as the background stars, and the 1st and 2nd order of the contaminant trace recorded in the detector. Similar to the \firefly reduction, the background is removed prior to the 1/f correction, and added back before the ramp fitting step. We perform the final background correction in Stage 2, at the integration level. We scale STScI's publicly available background model independently before and after the ``step" occurring around the 750th detector column. We use the upper portion of the detector, free from both the target's and contaminant's trace flux, to compute the scaling parameter. We correct for bad pixels and cosmic rays both spatially and temporally, and finally extract the stellar spectrum using a box extraction with a full-width aperture of 36 pixels, as it minimizes the white light out-of-transit scatter. During the spectral extraction step, the wavelength solution is refined by cross-correlating the extracted spectrum with a PHOENIX model consistent with LHS-1140's stellar parameters.

We perform all light curve fits using \texttt{juliet} \citep{Espinoza2019}, adopting the \texttt{batman} transit model \citep{Kreidberg2015} and the nested sampling algorithm \texttt{dynesty} \citep{Speagle2020} with 1000 and 500 live points for the broadband and spectroscopic light curve fits, respectively. For each visit, we first produce a broadband light curve by summing the flux from 0.85 to 2.8$\,\micron$, which we fit first in order to derive the system parameters.  We fix the period to 24.73723 days \citep{Cadieux2024}, and fit for the mid-transit time $T_0$, semi-major axis $a/R_s$, the impact parameter $b$, the planetary radius $R_p/R_s$, the baseline flux, a jitter term and a linear trend to account for systematics. We also fit for the limb darkening parameters $q_1$ and $q_2$, using the parameterization from \cite{Kipping2013}.
We then fit the spectral light curves at the pixel level around the helium triplet, from $1.07$ to $1.095\,\micron$. We fix the system parameters to the corresponding broadband light curve values, and fit only for the planetary radius $R_p/R_s$, the baseline flux, a jitter term and a linear trend. During the spectroscopic fits we fix the limb darkening parameters to the \texttt{PHOENIX} models \citep{Husser2013}, assuming the same stellar parameters as \firefly. As proposed in previous works \citep[e.g.][]{Espinoza2015, maxted2023}, we rescale the intensity profiles to account for the spherical geometry used in the \texttt{PHOENIX} models, and 
compute the corresponding limb darkening parameters. Furthermore, the broadband light curve of Visit 4 shows evidence for three possible spot-crossing events during the transit of \planetname. We mask the affected portions of the light curves during our broadband and spectroscopic fits, resulting in increased error bars in the derived transit depths when compared to the \firefly analysis, though the shape of the spectrum is unchanged regardless of whether we mask or ignore these points in the \exotedrf reduction. We show the resulting \exotedrf spectra compared to that of \firefly in \autoref{fig:reductions_compare}.

\subsection{Agreement Between Reductions}

\begin{figure*}
    \centering
    \includegraphics[width=1\linewidth]{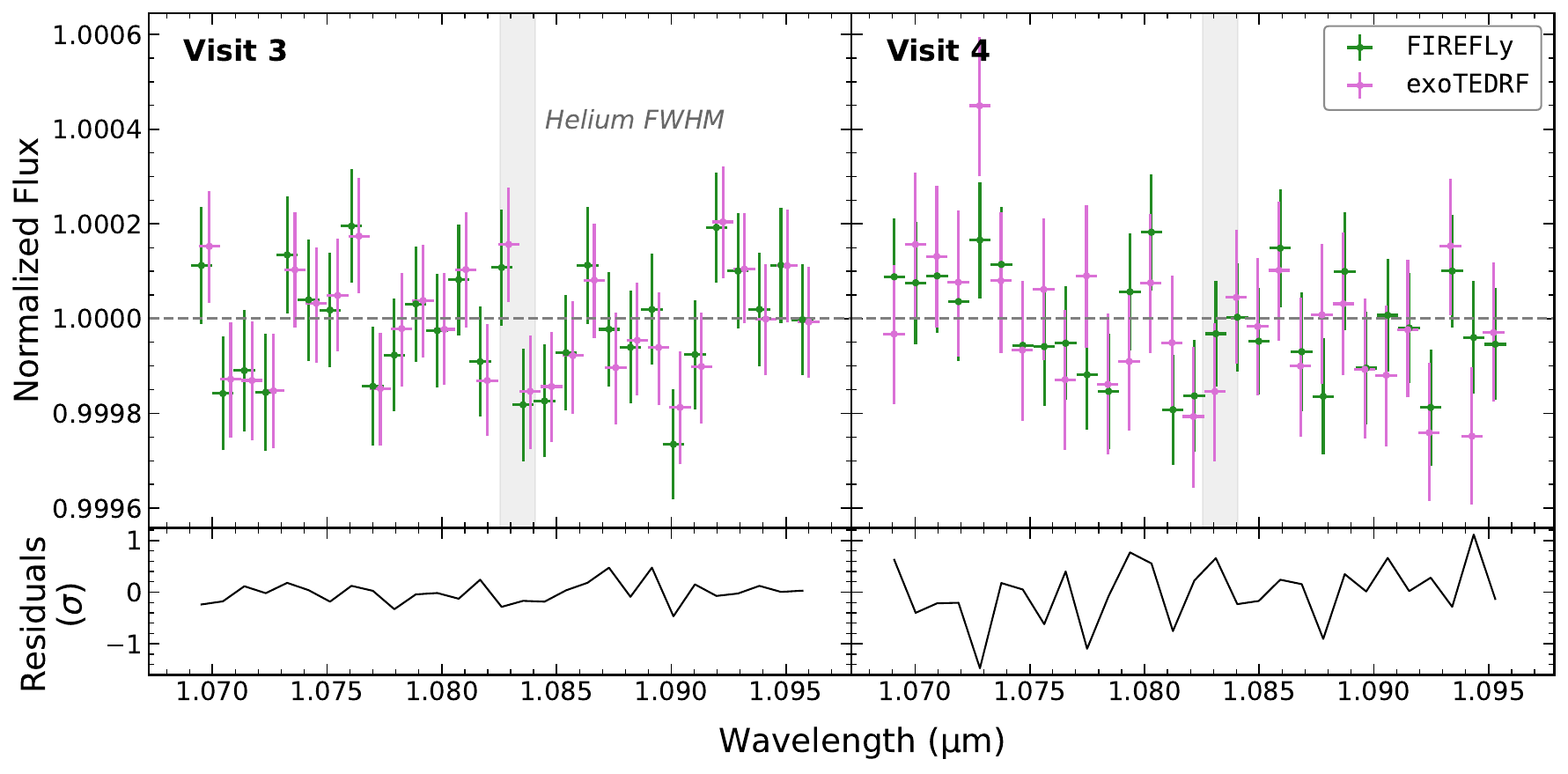}
    \caption{Comparison between \firefly and \exotedrf reductions for the newly acquired NIRISS/SOSS Visits 3 and 4 (from GO 7073). Top panels show the transmission spectra for the \firefly (green) and \exotedrf (pink) reductions. Bottom panels show the $\sigma$ differences between reductions, which are within $1\sigma$ for the vast majority of columns. The wavelength solution between reductions is inconsistent by 1/3 of a pixel for Visit 3, but the reductions are still in excellent agreement. For Visit 4, \exotedrf's error bars are somewhat larger than \firefly's due to \exotedrf's decision to mask the possible spot crossings in the transit, but the reductions are otherwise consistent. Both the \firefly and the \exotedrf reductions show no hint of helium absorption.}
    \label{fig:reductions_compare}
\end{figure*}

As seen in \autoref{fig:reductions_compare}, both reductions agree well, with no visual sign of helium in either reduction. For Visit 3, there is a slight wavelength offset in the central wavelength of each column between reductions, discussed in \autoref{sec:appendix_wvl_solutions}. Still, the shape of the transmission spectrum remains unchanged regardless of the wavelength offset. For Visit 4, the spectra are also consistent between our two pipelines. \exotedrf's error bars in Visit 4 are somewhat larger than \firefly's given \exotedrf's decision to mask the possible spots in the Visit 4 light curve, whereas \firefly ignores the spots and fits all the in-transit data points. Again, though, the agreement otherwise between the reductions demonstrates that our choice to mask or ignore the possible spots does not change our interpretation that no helium is detected in these spectra. 

\subsection{A Note on Wavelength Solutions} \label{sec:appendix_wvl_solutions}

The difference in wavelength solutions between \firefly and \exotedrf for Visit 3 may be due to \exotedrf's refinement of the wavelength solution using an M-dwarf model as a calibrator, whereas \firefly uses the standard \texttt{pastasoss} calibration. This latter technique predicts the wavelength solution using the x-positions of hydrogen stellar features, which have been calibrated using an A-star \citep{Baines2023a}. As stated in \cite{Baines2023a}, this technique provides an accurate wavelength solution down to a few tenths of a pixel. The \firefly wavelength solution is discrepant from \exotedrf's by a third of a pixel, so this falls within the expected uncertainty range of wavelength solutions for SOSS. 

For Visit 4, the wavelength agreement between reductions is better,
although we note that Visit 4's wavelength solution for \firefly is the most discrepant compared to the other three visits, by about 4/10ths of a pixel. This is not a concern, as the visit-to-visit wavelength solution is known to vary by a few pixels due to minor misalignments in the pupil wheel position (PWCPOS) at the time of the observation \citep{Baines2023a}. There is a correlation between pixel offset and PWCPOS (see Figure 3 in \citealt{Baines2023a}), and for PWCPOS values near the value recorded during Visit 4 (245.8276), pixel offsets relative to the \texttt{pastasoss} polynomial fit show scatter of up approximately a third of a pixel, consistent with our findings.

\bibliography{main}{}
\bibliographystyle{aasjournalv7}

\end{document}